# THE EISENSTEIN GROUP AND THE PSEUDO HYPERBOLIC FUNCTIONS

G. DATTOLI

ENEA - Dipartimento Tecnologie Fisiche e Nuovi Materiali
Centro Ricerche Frascati, Roma

M. MIGLIORATI

Dipartimento di Energetica
Università degli Studi di Roma "La Sapienza"
Via A. Scarpa, 14 - 00161 Rome, Italy

P. E. RICCI

Dipartimento di Matematica, Università di Roma "La Sapienza"
P.le Aldo Moro, 5 - 00185 Rome, Italy

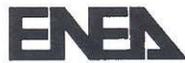

# THE EISENSTEIN GROUP AND THE PSEUDO HYPERBOLIC FUNCTIONS

G. DATTOLI

ENEA - Dipartimento Tecnologie Fisiche e Nuovi Materiali
Centro Ricerche Frascati, Roma

M. MIGLIORATI

Dipartimento di Energetica
Università degli Studi di Roma "La Sapienza"
Via A. Scarpa, 14 - 00161 Rome, Italy

P. E. RICCI

Dipartimento di Matematica, Università di Roma "La Sapienza"
P.le Aldo Moro, 5 - 00185 Rome, Italy

# THE EISENSTEIN GROUP AND THE PSEUDO HYPERBOLIC FUNCTIONS

G. DATTOLI, M. MIGLIORATI, P. E. RICCI

*Abstract*
*We review the theory of the pseudo-hyperbolic functions on the basis of an algebraic point of view which employs the Eisenstein group. We frame the theory within the general context of the number decomposition and discuss the importance of these functions in the theory of the generalized Fourier transforms.*

**Riassunto**
La teoria delle funzioni pseudo iperboliche viene formulata nel contesto di una teoria algebrica che utilizza il gruppo di Einstein.
Il problena della deefinizione di tali funzioni viene inquadrato nel contesto più generale della decomposizione dei numeri di Einstein. Si discute infine l'importanza delle funzioni pseudo iperboliche nell'ambito della teoria delle trasformate di Fourier generalizzata.

# INDICE

# THE EISENSTEIN GROUP AND THE PSEUDO HYPERBOLIC FUNCTIONS

## 1. INTRODUCTION

The Eisenstein integers form a commutative ring of algebraic integers in the algebraic number field $Q(\sqrt{-3})$ [1]. They are complex numbers of the form

$$z = x + \hat{\omega}\, y \tag{1}$$

with $\hat{\omega}$ being a Eisenstein unit, defined as

$$\hat{\omega} = -\frac{1 - i\sqrt{3}}{2} = e^{i\frac{2\pi}{3}} \tag{2}$$

The group of the Eisenstein units in the ring of the Eisenstein integers is the cyclic group, formed by the three roots of the unity in the complex plane. The elements of this group are

$$\{\pm 1, \pm \hat{\omega}\} \tag{3}$$

and it is easily checked that

$$\begin{aligned}\hat{\omega}^2 + \hat{\omega} &= -1, \\ \hat{\omega}^3 &= 1\end{aligned} \tag{4}.$$

The ring of the Eisenstein integers forms a Euclidean domain, whose norm is specified by

$$(a + \hat{\omega}\, b)(a + \hat{\omega}^2 b) = a^2 - a\, b + b^2 \tag{5}.$$

The above identity holds also for non integer values of $a, b$ and along with (5) it is worth mentioning that

$$a^3 + b^3 = (a+b)(a+\hat{\omega} b)(a+\hat{\omega}^2 b) \tag{6}$$

A fairly straightforward, but important consequence, of the cyclical properties of the Eisenstein units is the identity

$$e^{\hat{\omega} x} = e_0(x) + \hat{\omega}\, e_1(x) + \hat{\omega}^2 e_2(x),$$
$$e_m(x) = \sum_{n=0}^{\infty} \frac{x^{3n+m}}{(3n+m)!}, \qquad m=0,1,2 \tag{7}$$

The functions $e_m(x)$ are an example of pseudo-hyperbolic functions (PHF)[1], of the type introduced in ref. (2).

The use of the group property of the unit $\hat{\omega}$, yields the following identity

$$e_m(\hat{\omega} x) = \hat{\omega}^m e_m(x) \tag{8a}$$

specifying the symmetry properties of the PHF under reflection on the Eisenstein lattice, see Fig. 1.

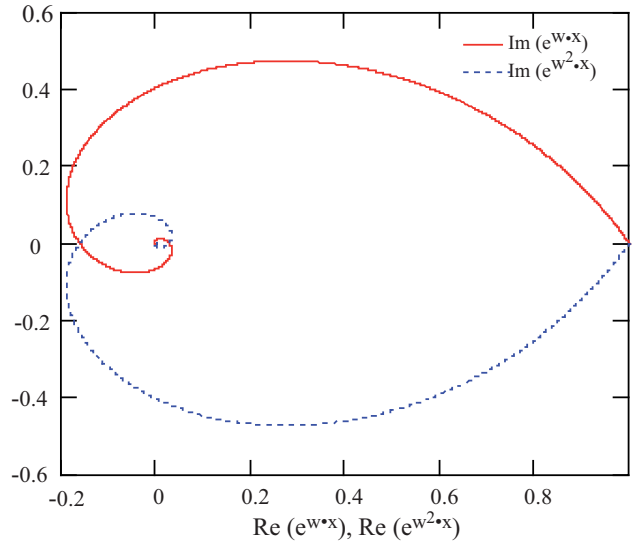

Fig. 1 - *Imaginary vs real parts of $e^{\hat{w} x}$ (continuous line) and of $e^{\hat{w}^2 x}$ (dot line)*

---

[1] These function could be denoted as $e_m(x;\omega), m=0,1,2$. Within this context we should denote with $e_m(x;1), m=0,1$ the ordinary Hyperbolic functions and with $e_m(x;i), m=0,1$ the ordinary circular functions. We use the definition given in eq. (7) to avoid a heavy notation.

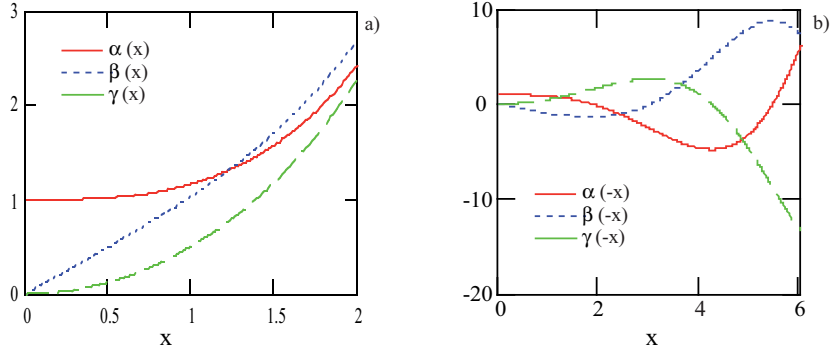

*Fig. 2 – a) The PHF vs. x. b) Reflected PHF vs. x*

In Figure 2 we provide a geometrical interpretation of the PHF along with the "evolution" in the Argand plane of the functions $e^{\hat{\omega} x}$, $e^{\hat{\omega}^2 x}$.

The modulus of $e^{\hat{\omega} x}$, $e^{\hat{\omega}^2 x}$ is not a constant as it follows from the fact that

$$e^{\hat{\omega} x} = e^{-\frac{x}{2}} e^{i\frac{\sqrt{3}}{2} x},$$
$$e^{\hat{\omega}^2 x} = e^{-\frac{x}{2}} e^{-i\frac{\sqrt{3}}{2} x}$$
(8b)

These introductory remarks have been aimed at defining and deriving the essential properties of the PHF, in the forthcoming section we will see how further consequences follow from the properties of the Eisenstein group.

## 2. THE ELEMENTARY PROPERTIES OF THE PHF

The use of the identities (7) and (4) can be exploited to write the PHF in terms of the complex exponentials as it follows

$$e_m(x) = \frac{1}{3} \sum_{j=0}^{2} \hat{\omega}^{(3-m)j} e^{\hat{\omega}^j x},$$
$$m = 0,1,2$$
(9)

furthermore, by keeping the derivative of both sides of eq. (9), we find the following properties under derivation

$$\frac{d}{dx} e_m(x) = e_{m-1}(x),$$
$$e_{-1}(x) \equiv e_2(x),$$
$$\left(\frac{d}{dx}\right)^3 e_m(x) = e_m(x), \qquad (10)$$
$$e_m(0) = \delta_{m,0}$$

The addition theorem for the PHF can be obtained by noting that

$$e_0(x+y) = e^{y\partial_x} e_0(x) = \sum_{r=0}^{\infty} \frac{y^r}{r!} \left(\frac{d}{dx}\right)^r e_0(x) =$$
$$= \left[ e_0(y\frac{d}{dx}) + e_1(y\frac{d}{dx}) + e_2(y\frac{d}{dx}) \right] e_0(x) \qquad (11)$$
$$= e_0(y)e_0(x) + e_1(y)e_2(x) + e_2(y)e_1(x)$$

The same procedure can be extended to the other functions, thus finding

$$e_m(x+y) = \sum_{j=0}^{2} e_{m+j}(y) e_{m-j}(x), \qquad (12).$$

The PHFs have not definite parity properties under the ordinary reflection $x \to -x$, the $e_m(-x)$ can however be expressed in terms of the $e_m(x)$ by setting $y = -x$ in eq. (12) and by noting that

$$\sum_{j=0}^{2} e_{m+j}(-x) e_{m-j}(x) = \delta_{m,0} \qquad (13).$$

Equation (13) is the fundamental identity of PHF[2], we can however use it as a system of algebraic equation in which $e_m(-x)$ are the unknown terms, thus finding

---

[2] In the case of ordinary HF we find
$$e_0(x;1) = \cosh(x), \; e_1(x;1) = \sinh(x),$$
$$e_0(-x;1) = e_0(x;1), \; e_1(-x;1) = -e_1(x;1)$$
therefore eq. (13) amounts to
$$e_0(-x;1)e_0(x;1) + e_1(-x;1)e_1(x;1) = \cosh(x)^2 - \sinh(x)^2 = 1$$

$$e_0(-x) = \frac{e_0(x)^2 - e_1(x)e_2(x)}{\Delta(x)},$$

$$e_1(-x) = \frac{e_2(x)^2 - e_0(x)e_1(x)}{\Delta(x)}, \quad (14).$$

$$e_2(-x) = \frac{e_1(x)^2 - e_0(x)e_2(x)}{\Delta(x)}$$

$$\Delta(x) = e_0(x)^3 + e_1(x)^3 + e_2(x)^3 - 3e_0(x)e_1(x)e_2(x)$$

The functions $e_m(x), e_m(-x)$ are reported in Fig. 2. They have totally different behaviours, the PHF are indeed similar to the ordinary hyperbolic functions, while their reflected counterparts exhibit an oscillatory behaviour.

Before concluding this section, let us note that if a given function $f(x)$ is specified by the series expansion $f(x) = \sum_{n=0}^{\infty} \frac{a_n}{n!} x^n$ then the following identity holds too

$$f(\hat{\omega} x) = \sum_{m=0}^{2} \hat{\omega}^m f_m(x),$$

$$f_m(x) = \sum_{n=0}^{\infty} \frac{a_{3n+m}}{(3n+m)!} x^{3n+m} \quad (15)$$

which, in analogy with eq. (9), yields the further identity

$$f_m(x) = \frac{1}{3} \sum_{j=0}^{2} \hat{\omega}^{(3-m)j} f(\hat{\omega}^j x) \quad (16).$$

Equations (15) and (16) should be considered within the more general framework of the theory of functions with complex variables, this topic goes beyond the scope of this paper and will not be further discussed[3].

---

[3] We can indeed define the variable
$z = x + \hat{\omega} y + \hat{\omega}^2 z,$
thus getting
$f(z) = u(x,y,z) + \hat{\omega} v(x,y,z) + \hat{\omega}^2 w(x,y,z)$
the analogous of the Cauchy-Riemann conditions, linking the derivatives of the $u, v, w$, can easily be stated.

## 3.   SPECIAL POLYNOMIALS AND THE EISENSTEIN UNITS

The importance of the PHF for the theory of special polynomials has been considered in Refs. [3,4], here we will discuss the problem from a different perspective.

We start our discussion with the following identity,

$$(x + \hat{\omega} y)^n = \sum_{j=0}^{2} \hat{\omega}^j h_n^{(3)}(x,y;j),$$

$$h_n^{(3)}(x,y;j) = n! \sum_{r=0}^{[n-j/3]} \frac{x^{n-3r-j} y^{3r+j}}{(n-j-3r)!(3r+j)!}$$

(17)

whose proof can easily be achieved using the generating function method, and $h_n^{(3)}(x,y;j)$ are the pseudo Hermite polynomials of the type introduced in Ref. [3].

The use of this last identity allows the definition of a "new" family of Laguerre polynomials.

The use of the formalism of the negative derivative operator allows to define the Laguerre polynomials as [5]

$$L_n(x,y) = (y - \hat{D}_x^{-1})^n,$$

$$L_n(x,y) = n! \sum_{r=0}^{n} \frac{(-1)^r y^{n-r} x^r}{(n-r)!(r!)^2}$$

(18)

where the operator $\hat{D}_x^{-1}$ is defined in such a way that [6][4]

$$\hat{D}_x^{-n} 1 = \frac{x^n}{n!}$$

(19).

Replacing $\hat{D}_x^{-1} \to \hat{\omega} \hat{D}_x^{-1}$ in the first of eq. (18), we obtain

---

[4]  The unit 1 appearing on the r h s of eq. (19) constitutes a kind of "vacuum state" where operators like the negative derivative acts, it should appear on r h s of eq. (18) too, but it has been omitted for the sake of conciseness.

$$(y - \hat{\omega}\hat{D}_x^{-1})^n = \sum_{j=0}^{2} \hat{\omega}^j l_n^{(3)}(y,x;j),$$

$$l_n^{(3)}(y,x;j) = h_n^{(3)}(y,\hat{D}_x^{-1};j) = n! \sum_{r=0}^{[n-j/3]} \frac{y^{n-3r-j} x^{3r+j} (-1)^{3r+j}}{(n-3r-j)! [(3r+j)!]^2}$$

(20).

The polynomials $l_n^{(3)}(.,.,.)$ define a kind of hybrid family (in between Hermite and Laguerre) of the type studied in Ref. (7).

The combination of operational methods and of the properties of the Eisenstein group offers further possibilities. We note indeed that, being

$$e^{y\partial_x^2} x^n = H_n(x,y),$$

$$H_n(x,y) = n! \sum_{r=0}^{[n/2]} \frac{x^{n-2r} y^r}{(n-2r)! r!}$$

(21)

we find

$$e^{\hat{\omega} y \partial_x^2} x^n = H_n(x, \hat{\omega} y) = \sum_{j=0}^{2} \hat{\omega}^j \eta_n(x,y;j),$$

$$\eta_n(x,y;j) = n! \sum_{r=0}^{[n/2]} \frac{x^{n-2(3r+j)} y^{3r+j}}{(n-2(3r+j))! (3r+j)!}$$

(22).

Furthermore the use of the generating function

$$e^{xt+yt^2} = \sum_{n=0}^{\infty} \frac{t^n}{n!} H_n(x,y) \tag{23}$$

can be exploited to get

$$e^{\hat{\omega} x + \hat{\omega}^2 y} = \sum_{j=0}^{2} \hat{\omega}^j g_j(x,y),$$

$$g_j(x,y) = \sum_{n=0}^{\infty} \frac{H_{3n+j}(x,y)}{(3n+j)!}$$

(24)

These last formulae have been derived in refs. [8], within the framework of a different formalism and have been exploited in the theory of arbitrary order coherent states [9].

It is worth noting that, in analogy to eq. (9), we find

$$g_m(x,y) = \frac{1}{3}\sum_{j=0}^{2}\hat{\omega}^{(m-3)j}e^{\hat{\omega}^j x + \hat{\omega}^{2j}y} \qquad (25a),$$

which yields e. g. the explicit forms

$$g_0(x,y) = \frac{1}{3}\left\{\left[1 + 2e^{\frac{1}{2}(x+y)}\cos(\frac{\sqrt{3}}{2}(x-y))\right]\cosh(x+y) \right.$$
$$\left. + \left[1 - 2e^{\frac{1}{2}(x+y)}\cos(\frac{\sqrt{3}}{2}(x-y))\right]\sinh(x+y)\right\} \qquad (25b).$$

Analogous relations for the Laguerre polynomials will be discussed in the concluding remarks.

## 4. PHF AND ASSOCIATED FOURIER TRANSFORM

The results of the previous sections suggest that the PHF and the cyclic properties of the Eisenstein group provide a tool to go beyond the ordinary concept of parity, accordingly we will say that a given function exhibits a degree of parity m on the Eisenstein lattice if it satisfies the identity

$$g(\hat{\omega}^m x) = \hat{\omega}^m g(x) \qquad (26).$$

It is therefore evident that any function can be decomposed in its m-parity components on the basis of eq. (16). The PHF are therefore the components with parity of order 3 of the exponential function.

The ordinary parity can be identified as a parity of order 2 and therefore the hyperbolic functions are the components with order 2 parity of the exponential function.

Such a concept can usefully be exploited to define the following Eisenstein-Fourier transform (EFT)

$$\tilde{f}(k;\hat{\omega}) = \frac{1}{\sqrt{2\pi}} \int_{-\infty}^{\infty} f(x) e^{-k\hat{\omega}x} dx \tag{27a}$$

which can also be cast in the form

$$\tilde{f}(k;\hat{\omega}) = \sum_{m=0}^{2} \hat{\omega}^m \tilde{f}_m(k),$$

$$\tilde{f}_m(k) = \frac{1}{\sqrt{2\pi}} \int_{-\infty}^{\infty} f(x) e_m(-kx) dx \tag{27b}.$$

It is evident that $\tilde{f}_m(k)$ plays the same role as sin and cos components of the ordinary FT.

The condition under which a given function admits an EFT can be easily stated, since the definition (27b) can be reduced to

$$\tilde{f}(k;\hat{\omega}) = \frac{1}{\sqrt{2\pi}} \int_{-\infty}^{\infty} \left[ f(x) e^{\frac{1}{2}kx} \right] e^{-i\frac{\sqrt{3}}{2}kx} dx \tag{28}$$

and therefore its existence is limited to those functions $f(x) e^{(1/2)kx}$ admitting an ordinary FT for some k-interval.

The Gaussian function admits an EFT whose real and imaginary parts are reported in Fig. 3.

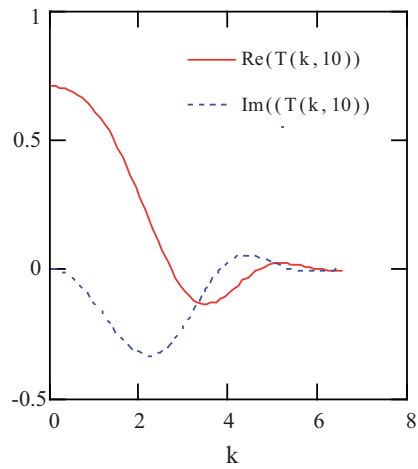

*Fig. 3 – Real (continuous line) and imaginary (dotline) parts of the Einstein Fourier transform of a gaussian*

It is interesting to note that, unlike the ordinary FT the EFT of the Gaussian satisfies the following properties

$$\frac{1}{\sqrt{2\pi}}\int_{-\infty}^{\infty}e^{-x^2}e_0(kx)dx = \frac{1}{\sqrt{2}}e_0(\frac{k^2}{4}),$$

$$\frac{1}{\sqrt{2\pi}}\int_{-\infty}^{\infty}e^{-x^2}e_1(kx)dx = \frac{1}{\sqrt{2}}e_2(\frac{k^2}{4}), \quad (29).$$

$$\frac{1}{\sqrt{2\pi}}\int_{-\infty}^{\infty}e^{-x^2}e_2(kx)dx = \frac{1}{\sqrt{2}}e_1(\frac{k^2}{4})$$

Before closing this section it is worth emphasizing that the following further form of EFT can be introduced

$$\tilde{f}(k;\hat{\omega}^2) = \frac{1}{\sqrt{2\pi}}\int_{-\infty}^{\infty}f(x)e^{-k\hat{\omega}^2 x}dx = \frac{1}{\sqrt{2\pi}}\int_{-\infty}^{\infty}\left[f(x)e^{\frac{1}{2}kx}\right]e^{i\frac{\sqrt{3}}{2}kx}dx \quad (30)$$

Such a definition is in some sense complementary to the previous form (28) and can be combined with it to get

$$\tilde{f}_c = \frac{\tilde{f}(k;\hat{\omega}^2) + \tilde{f}(k;\hat{\omega})}{2},$$

$$\tilde{f}_s = \frac{\tilde{f}(k;\hat{\omega}^2) - \tilde{f}(k;\hat{\omega})}{2i} \quad (31).$$

In a forthcoming dedicated paper we will discuss how most of the theorem valid for the ordinary FT can be extended to the EFT.

## 5. CONCLUDING REMARKS

In the introductory section we have pointed out that the functions $e_m(x)$ provide an extension of the hyperbolic functions, it might be therefore interesting to pursue the analogy by defining the following forms of hyperbolic tangents

$$t_{m,n}(x) = \frac{e_m(x)}{e_n(x)} \quad (32)$$

whose behaviour is shown in Fig. 4.

The derivative of the Eisenstein tangents reads

$$\frac{d}{dx} t_{m,n}(x) = t_{m-1,n}(x) - t_{n-1,n}(x) t_{m,n}(x) \tag{33}.$$

Furthermore, according to the reflection properties given in eq. (14), we find

$$t_{0,2}(-x) = \frac{t_{0,1}(x) - t_{2,0}(x)}{t_{1,0}(x) - t_{2,1}(x)} \tag{34}.$$

A function which deserves attention is the following extension of the hyperbolic secant (see Fig. 5, for a comparison with the ordinary hyperbolic secant)

$$s(x) = \frac{1}{e_0(x)} \tag{35}$$

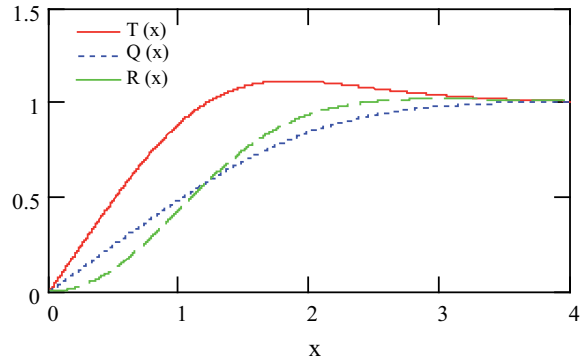

*Fig. 4 - Eisenstein tangents: $t_{1,0}(x)$ continuous line $t_{2,1}(x)$ dot line, $t_{2,0}(x)$ dash line*

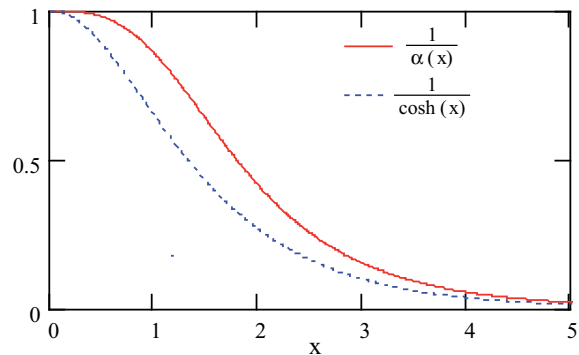

*Fig. 5 - Eisenstein secant (continuous line) and ordinary hyperbolic secant (dot line)*

satisfying the differential equation

$$s'''(x) + 3t_{2,0}(x)s''(x) + 3t_{1,0}(x)s'(x) + s(x) = 0 \qquad (36).$$

The Euclidean norm associated with the Eisenstein group can be used in different applicative contexts, here we will touch on an application in quantum mechanics, involving the following harmonic oscillator Hamiltonian

$$\hat{H} = \frac{1}{2}\hat{p}^2 - \frac{1}{4}(\hat{p}\hat{q} + \hat{q}\hat{p}) + \frac{1}{2}\hat{q}^2 \qquad (37)$$

where $\hat{p},\hat{q}$ are momentum position operators satisfying the commutation bracket

$$[\hat{p},\hat{q}] = -i \qquad (38)$$

According to eq. (5) we can recast the above Hamiltonian in the form[5]

$$\begin{aligned} H &= \hat{A}\hat{A}_c + \frac{i}{4}(2\hat{\omega} + 1), \\ \hat{A} &= \frac{\hat{q} + \hat{\omega}\hat{p}}{\sqrt{2}}, \\ \hat{A}_c &= \frac{\hat{q} + \hat{\omega}^2\hat{p}}{\sqrt{2}} \end{aligned} \qquad (39)$$

The operators $\hat{A}, \hat{A}_c$ satisfy the commutation relations

$$[\hat{A},\hat{A}_c] = -\frac{i}{2}(2\hat{\omega} + 1) \qquad (40)$$

and play a role analogous to that of creation and annihilation operators.

The Eisenstein coherent states can be introduced in terms of the ordinary creation operator according to the following definition

$$|\alpha;\hat{\omega}\rangle = e^{\hat{\omega}\alpha\hat{a}}|0\rangle \qquad (41)$$

where $|0\rangle$ is the harmonic oscillator vacuum state.

According to the previously discussed properties of the Eisenstein unit, we obtain

$$|\alpha;\omega> = \frac{1}{3}\sum_{j=0}^{2}\hat{\omega}^{j}|\alpha_j>,$$

$$|\alpha_j> = \sum_{n=0}^{\infty}\frac{\alpha^{3n+j}}{\sqrt{(3n+j)!}}|3n+j> \qquad (42)$$

The Eisenstein components of the coherent field $|\alpha>_j$ are also easily shown to be eigen-states of the operator $\hat{a}^j$, we find indeed

$$\hat{a}^j|\alpha>_j = \alpha^j|\alpha>_j \qquad (43)$$

The properties of analogous type of quantum states have been discussed in refs [9], further comments on their properties and applications will be presented elsewhere.

Before concluding this paper let us mention that the PHF can be extended by exploiting the set of units of the Kummer ring [1] defined by[6]

$$\left\{,\hat{\omega}_{m,1},\mathrm{K},\hat{\omega}_{m,m-1}\right\}$$

$$\hat{\omega}_{m,k} = \hat{\omega}_m^k = e^{\frac{ik\pi}{m}}, 1 \le k \le m-1 \qquad (44).$$

---

[5] It is worth noting that the operator $\hat{A}\hat{A}_c$ is not a Hermitian, it cannot be therefore considered a number operator. The consequence of this fact will be discussed in a forthcoming paper.

[6] Note that in analogy to eq. (6) we also find
$$a^n + b^n = \prod_{r=0}^{n-1}(a+\hat{\omega}_n^r b)$$

Even though these functions have been defined in [2] their reformulation in algebraic terms may offer further elements of generality.

A fairly straightforward application of this extended family of PHF is provided by the following example

$$e^{\sum_{j=1}^{m-1} \hat{\omega}_{m,j} x_j} = \sum_{n=0}^{\infty} \frac{\hat{\omega}_{m,n}}{n!} H_n^{(m-1)}(\{x\}_1^{m-1}) = \sum_{j=0}^{m-1} \hat{\omega}_{m,j} g_j(\{x\}_1^{m-1}),$$

$$g_j(\{x\}_1^{m-1}) = \sum_{n=0}^{\infty} \frac{1}{(3n+j)!} H_{3n+j}^{(m-1)}(\{x\}_1^{m-1}) \qquad (45).$$

$$H_n^{(p)}(\{x\}_1^p) = n! \sum_{r=0}^{[n/p]} \frac{x_p^r H_{n-pr}^{(p-1)}(\{x\}_1^{p-1})}{(n-pr)! r!}$$

Further applications will be discussed in a forthcoming investigations where we will treat more deeply the topics we have just touched here.